\documentclass[aps,prb, twocolumn,superscriptaddress, 
                                        showpacs,amsmath]{revtex4}
\usepackage{bm}
\usepackage{graphicx}

\begin{document}
\title{Numerical simulation evidence of spectrum rearrangement 
in impure graphene}

\author{S.S. Pershoguba}
\affiliation{Moscow Institute of Physics and Technology, 
             Institutskii per. 9, Dolgoprudny 141700, 
             Moscow Region, Russia}
\affiliation{G. V. Kurdyumov Institute of Metal Physics, 
             National Academy of Sciences of Ukraine
             Vernadsky Ave. 36,
             Kyiv 03680, Ukraine}
\author{Yu.V. Skrypnyk}
\affiliation{G. V. Kurdyumov Institute of Metal Physics, 
             National Academy of Sciences of Ukraine
             Vernadsky Ave. 36,
             Kyiv 03680, Ukraine}
\author{V.M. Loktev}
\affiliation{Bogolyubov Institute for Theoretical Physics,
             National Academy of Sciences of Ukraine
             Metrolohichna Str. 14-b,
             Kyiv 03680, Ukraine}

\begin{abstract}
By means of numerical simulation we confirm that in graphene with
point defects a quasigap opens in the vicinity of the resonance state
with increasing impurity concentration. We prove that states inside
this quasigap cannot longer be described by a wavevector and are
strongly localized. We visualize states corresponding to the density
of states maxima within the quasigap and show that they are yielded by
impurity pair clusters.
\end{abstract}



\maketitle

\section {Introduction}

Not so long ago, graphene has been cleaved out for the first time by
the so--called scotch--tape technique.\cite{Novo} Since that, this
novel, truly two--dimensional material continues to put new challenges
to scientific community. Graphene is already known to manifest some
remarkable properties. The most unusual of them, and, correspondingly,
the most frequently emphasized on, is the Dirac dispersion of Fermi
elementary excitations. This unusual spectrum makes graphene a rather
promising material for a variety of applications in computer
electronics and chemical sensors. While graphene is known to possess
outstanding structural stability and crystalline quality, existing
methods of its isolation necessarily imply the presence of a certain
amount of defects. Moreover, impurities can be embedded into graphene
intentionally in order to tune up its physical characteristics in  
accordance with a specific application. Even though some applications
are destined for a distant future, the need in deliberate and proper
functionalization of graphene provides adequate grounds for an
extensive study of different types of defects in this
material. Despite a noticeable quantity of papers devoted to the study of
impure graphene, a comprehensive understanding of impurity effects on
its electron spectrum is still lacking.  

While different  types of disorder are inherent in graphene, below
we are particularly interested among them in the substitutional point
defects. This  commonly used model applies not only to chemically
substituted carbon atoms  or an absence of them (vacancies), but also,
to a known extent, to adsorbed atoms, molecules, or radicals on the
graphene sheet.\cite{adatoms} Regarding the substitutional
impurities in graphene, comprehensive attention has been paid as to
the single impurity problem, in which the impurity state wave function
has been studied for a single defect and a pair of
them,\cite{Katsnelson} as to the evolution of the density of states
(DOS) in graphene with increasing the impurity
concentration.\cite{Pereira1,Pereira2,Sarma,Pereira,Wu} However, such
a phenomenon as spectrum rearrangement is frequently overlooked. The
main concept of the spectrum rearrangement is based on the existence
of some critical impurity concentration, at which the spectrum of a
disordered system undergoes a cardinal qualitative change. As a rule,
the spectrum rearrangement should be related to the appearance of a
local level or a resonance state. Characteristics of this impurity
state, namely its energy and damping, which are determined within the
single impurity problem, are shaping the scenario and type of the
subsequent spectrum rearrangement. Albeit a single point defect is
expected to perturb only the lattice cite occupied by an impurity or,
at most, the adjacent lattice cites, and thus should be classified as
a short-range defect, the effective radius of the correspondent
impurity state can far exceed the lattice constant, when its energy is
close to the van Hove singularity in the spectrum. Spacial overlap of
individual impurity states, which occurs with increasing the impurity
concentration, is marking the onset of the spectrum rearrangement. This
simple consideration gives a possibility to roughly estimate the
critical concentration of the spectrum rearrangement. Since the
effective radius of the single impurity state in certain cases can be
large compared to the lattice constant, the respective critical
concentration should be fairly low. As a result, in such situations
only a trace amount of impurities can provoke the spectrum
rearrangement.

In graphene, the spectrum rearrangement driven by an increase in
the concentration of defects described by the Lifshitz model has been
analytically examined in Refs.~\onlinecite{YuV,YuV0}. It has been
demonstrated that the passage of the spectrum rearrangement is of the
anomalous type due to a weak resonance state. That means that for a
sufficiently large compared to the bandwidth impurity potential a
quasigap is gradually opening around the resonance energy and at the
critical concentration spans up to the Dirac point in the
spectrum. With further increase in the impurity concentration
(i.e. after the spectrum rearrangement), the quasigap is rapidly
enlarging. In this regime the width of the quasigap is approximately
proportional to the square root of the impurity concentration. In
Refs.~\onlinecite{YuV,YuV0} the analysis of the spectrum rearrangement
in graphene has been conducted by means of the coherent potential
approximation (CPA) applicability criterion, which is instrumental in
determining spectrum domains with different degree of
localization.\cite{sprb} 

The aim of the current work is to carry out the numerical simulation
of graphene DOS with the special emphasis on the spectrum
rearrangement phenomenon (for its physical description see also the
review Ref.~\onlinecite{ilp}). By considering for each chosen
perturbation strength those impurity concentrations that are close to
the expected critical concentration of the spectrum rearrangement, we
show that implementing the criterion of the CPA applicability it is
possible to judge upon the spectrum rearrangement process in
graphene. As a next step in our previous attempts,\cite{YuV,YuV0} we
are paying a special attention to the CPA validity. By comparing the
CPA output with the numerical results, we verify that the CPA
applicability criterion works rather well. We also notice that when
impurity concentration is low enough, the average T-matrix
approximation (ATA) is in good agreement with numerically calculated
DOS. We discuss the spectrum rearrangement in graphene and the
correspondent interplay between numerical and analytical
results. After the spectrum rearrangement takes place, we identify the
cluster structure of graphene's DOS in the vicinity of the impurity
resonance energy. The structure observed evidently cannot be explained
by means of available analytic approaches and requires further
analysis.

As an effective tool for numerical calculations we employ the negative
eigenvalue theorem as suggested by Dean.\cite{Dean} This approach, to
the best of our knowledge, hasn't been used for graphene yet (see,
e.g. Refs.~\onlinecite{Pereira,Wu}). Its advantages are discussed
below. Finally, we describe a drift of the Fermi energy from the DOS
minimum, a kind of a self--doping effect, when Fermi level shifts away
without actual introduction of additional carriers into the disordered
system. 

The paper is organized as follows: in Section II A we remind the
basic mathematical impurity model. In Section II B we introduce the
concept of spectrum rearrangement. In Section II C we briefly set
forth the numerical approach. In Section III we present and discuss
results. In the last Section we summarize outcomes of our study.

\section {Model disordered system and methods of its analysis}

\subsection{Impurity model}

In the tight--binding approximation the simplest (for spinless
fermions) graphene Hamiltonian has the form,
\begin{equation}
   \mathbf{H_0} = t \sum\limits_{<\bm{n}_\alpha, \bm{m}_\beta>}%
   \mathbf{a}_{\bm{n}_\alpha}^\dag \mathbf{a}^{}_{\bm{m}_\beta}.
\label{hamiltonian}
\end{equation}
Here, $t\approx 2.7 eV$ is the hopping integral, $\bm{n}$ and $\bm{m}$
denote vectors of lattice cells, Greek indices $\alpha$ and $\beta$
correspond to the graphene sublattices, and summation runs over all
nearest neighbors. It has been confirmed experimentally that this
approximation describes graphene's electronic spectrum fairly
well.\cite{Bostwick} 

\begin{figure}
\includegraphics[width=0.475\textwidth]{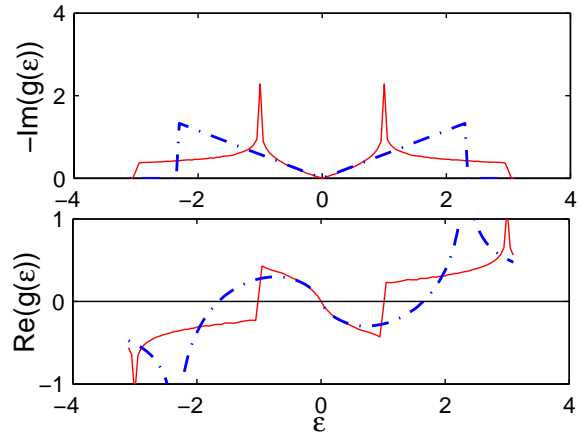}
\caption{(color online) A comparison between the exact $g_0(\epsilon)$
(solid line) and its analytical approximation (\ref{gf0}) (dash-dotted
line).} 
\label{comparison}
\end{figure}

The diagonal element of the Green's function in the cite
representation,
\begin{equation} 
g_{\bm{n}_\alpha \bm{m}_\beta}(\varepsilon)=\lim_{\delta\rightarrow 0^{+}}%
\sum\limits_{j}\frac{\psi^{}_{\bm{n}_\alpha}(j)\psi^{\dag}_{\bm{m}_\beta}(j)}%
{\varepsilon+i\delta-\varepsilon(j)},
\end{equation}
where the summation runs over all eigenstates $\psi(j)$, and
$\varepsilon(j)$ are their corresponding eigenvalues, in the case of
Hamiltonian (\ref{hamiltonian}) can be easily approximated by
\begin{equation}
    g_0(\varepsilon) = g_{0_\alpha 0_\alpha}(\varepsilon)  \approx
\frac{2\varepsilon}{W^2} ln \left(\frac{|\varepsilon |}{W}\right)-
i\pi\frac{|\varepsilon|}{W^2} \label{gf0}
\end{equation}
in the low--energy limit, i.e.  $|\varepsilon|\ll W =%
\sqrt{\pi\sqrt{3}}t$, where $W$ is the bandwidth. A detailed
derivation of (\ref{gf0}) can be found in Refs.~\onlinecite{YuV,YuV0}.
A comparison between the exact diagonal element of the Green's function
and its low--energy approximation is given in Fig.~\ref{comparison}.
Since we are interested only in the relatively close vicinity of the
Dirac point in the spectrum, the approximation (\ref{gf0}) looks
appropriately shaped.
 
While the host DOS can be straightforwardly found from the imaginary
part of the latter expression, impurities break the translational
symmetry and so the DOS of disordered graphene (which is the focus of
the current investigation) cannot be directly obtained. Point defects
in graphene are usually modeled by adding the following perturbation
in the Hamiltonian (the so--called Lifshitz model):
\begin{equation}
\mathbf {U} = V_L
\sum\limits_{\bm{p}_\alpha}\mathbf{a}_{\bm{p}\alpha}%
^\dag\mathbf{a}_{\bm{p}\alpha}^{},
\label{perturbation}
\end{equation}
where $V_L$ is the impurity potential, and $\bm{p}_\alpha$ runs over
sites occupied by impurities. It is supposed that impurities are
distributed among lattice sites absolutely at random with
concentration $c$ representing the probability that an arbitrary site
is occupied by an impurity. Thus, for a large system with $N$ lattice
sites the total amount of impurities tends to $cN$. 

\subsection{Spectrum rearrangement and CPA applicability criterion}

When the impurity concentration is small enough, conventional analytic
approaches\cite{Elliot,Lifhitz} can be applied. To be concise, in the
first approximation the averaged perturbed Green's function
\begin{equation}
\mathbf{G}(\varepsilon)=<(\varepsilon-\mathbf{H_0-U})^{-1}> 
\end{equation}
can be found as some renormalization of the host Green's function:
\begin{equation}
\mathbf{G}(\varepsilon) = \mathbf{g}(\varepsilon-\sigma(\varepsilon))
\end{equation} 
Two cases are of a particular interest: the average T-matrix
approximation (ATA),
\begin{equation}
\sigma_{ATA} (\varepsilon) = \frac{cV_L}{1-(1-c)V_L g_0(\varepsilon)},
\label{ATA}
\end{equation}
which accounts for multiple single--site scattering by an impurity,
and the coherent potential approximation,
\begin{equation}
\sigma_{CPA}(\varepsilon)=\frac{cV_L}%
{1-[V_L-\sigma_{CPA}(\varepsilon)]%
g_0(\varepsilon-\sigma_{CPA}(\varepsilon))},
\label{CPA}
\end{equation}
which adds the self--consistency. In the CPA the self--energy is
taken from the requirement that the single--site renormalized T-matrix
should be zero on average. In both methods scatterings on pairs and
larger groups of impurities is omitted. Thus, these approximations are
expected to remain valid, when cluster effects are insignificant in a
disordered system. 

However, when impurity concentration is gradually increased,
individual impurity states (visualized for graphene, e.g., in
Ref.~\onlinecite{Katsnelson}) begin to overlap with each other. Thus,
a contribution from scatterings on impurity clusters to the
self--energy is becoming more pronounced in the vicinity of the
impurity state energy. As a result, a significant overlap between
impurity states corresponds to the commencement of substantial
modifications in the spectrum of a disordered system. In other words,
it points out the critical concentration of the spectrum
rearrangement. This simple reasoning provides a possibility to
estimate the critical concentration in graphene with Lifshitz
impurities. From the expression for the non-diagonal element of the
host Green's function,\cite{Bena} it can be deduced that an effective
decay radius of the impurity state is $r_{imp}\approx |V_L|$. It
should be noticed that in commonly encountered cases for the Lifshitz
impurity model, an increase in the parameter $V_L$ leads to the
intensification of the impurity state localization and, consequently,
to a decrease in $r_{imp}$. The opposite result for graphene is caused
by the particle--hole symmetry of the Dirac Hamiltonian. Another
characteristic space interval is the average distance between
impurities, which depends on impurity concentration as $<r>\sim
1/\sqrt{c}$. Both radii coincide ($<r>\approx r_{imp}$) at some
impurity concentration $c_r\sim 1/V_L^2$. Thus, the condition
$<r>\approx r_{imp}$ defines the spectrum rearrangement concentration
$c_r$. As has been argued in Ref.~\onlinecite{YuV}, at this critical
concentration a quasigap filled with strongly localized states should
sweep from the resonance energy to the Dirac point, stimulating an
accumulation of considerable changes in the DOS. Albeit this condition
is reasonably intuitive, it is too rough for an accurate forecast of
the critical concentration. Moreover, it does not provide any
information on the energy intervals, in which the spectrum is more
exposed to the rearrangement process. Since cluster effects, which
make up the core of the spectrum rearrangement,  are not included into
the CPA, the CPA applicability criterion can be successfully employed
for the analysis of the spectrum rearrangement process. This very
approach has been in fact realized in Refs.~\onlinecite{YuV,YuV0} for
the impure graphene. Namely, by following the conventional technique
of the Green's function cluster expansion\cite{ilp,Lifhitz} it is
possible to represent the self--energy as a series in all possible
groups of impurity centers. At that, the first term of this series
corresponds to the conventional CPA. The small parameter of the
series, 
\begin{multline}
 R(\varepsilon) =\\
=\left|c\left[\frac{V_L-\sigma_{CPA}(\varepsilon)}%
{1 -[V_L-\sigma_{CPA}(\varepsilon)] g_0(\varepsilon-\sigma_{CPA}%
(\varepsilon))}\right]^2+\right.\\
\left. +(1-c)\left[\frac{-\sigma_{CPA}(\varepsilon)}%
{1+\sigma_{CPA}(\varepsilon)g_0(\varepsilon-\sigma_{CPA}(\varepsilon))}%
\right]^2\right| \times \\
\times\left[\sum_{\bm{n}\neq 0}|g_{\bm{0}_1 \bm{n}_1}(\varepsilon-%
\sigma_{CPA}(\varepsilon))|^2+\right. \\
+\left.\sum_{\bm{n}}|g_{\bm{0}_1 \bm{n}_2}(\varepsilon-%
\sigma_{CPA}(\varepsilon))|^2\right],
\label{criterium0}
\end{multline}
is indicative of the relative strength of cluster effects at a given
energy, and can be used to outline qualitatively different spectrum
domains. 

In those spectral domains, where the small parameter of the series
$R(\epsilon)$ is high, the CPA is not reliable and correspondent
states are showing a tendency towards localization. For ordinary 3D
systems a mobility edge should be expected close to the energy, at
which $R(\epsilon)=0.5$.\cite{sprb} At the same time, it can be
demonstrated that the maximum magnitude of $R(\epsilon)$ is unity, and
it is reached at the van Hove singularities of the spectrum. At the
low impurity concentration, expression (\ref{criterium0}) can be
approximated by     
\begin{equation}
R(\varepsilon)=\left|\frac{\sigma_{CPA}^2(\varepsilon)}{c}%
\left[g_0(E)+\frac{\partial g_0(E)}{\partial E}\right]%
_{E=\varepsilon-\sigma(\varepsilon)}\right| \label{criterium}
\end{equation}
There are two factors in (\ref{criterium}). The first of them,
$\sigma_{CPA}^2(\varepsilon)/c$,  increases in absolute magnitude
around the impurity state energy, which can be determined from the
Lifshitz equation, $1=V_Lg_{0}(\varepsilon)$). The second one (in the
square brackets) is the sum of the Green's function and its derivative,
which increases in the vicinity of the Dirac point (or any other van
Hove singularity). Consequently, the energy dependence of the CPA
applicability criterion should possess different maxima, around which
the CPA is not valid. Even though the CPA applicability criterion has
been deduced for the fictitious system with a single Dirac cone in the
spectrum, it will be apparent below that it is an adequate tool for
the spectrum rearrangement analysis in the actual graphene. As regards
the CPA and the ATA, it is not difficult to show that presence of the
two different Dirac cones in graphene alters their output only in a
trivial way.

\begin{figure}
\includegraphics[width=0.475\textwidth]{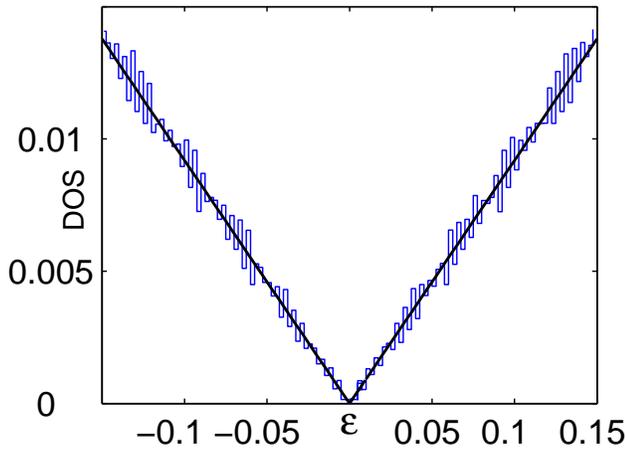}
\caption{(color online) Density of states for graphene without
impurities obtained by Dean's numerical method compared to the exact
one for the infinite system.} 
\label{ideal}
\end{figure}

\begin{figure}
\begin{tabular}{c}
\includegraphics[width=0.475\textwidth]{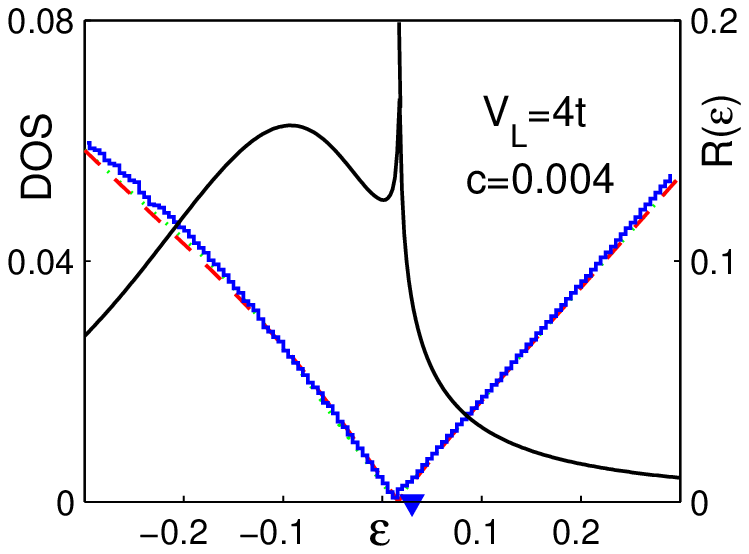}\\ 
\includegraphics[width=0.475\textwidth]{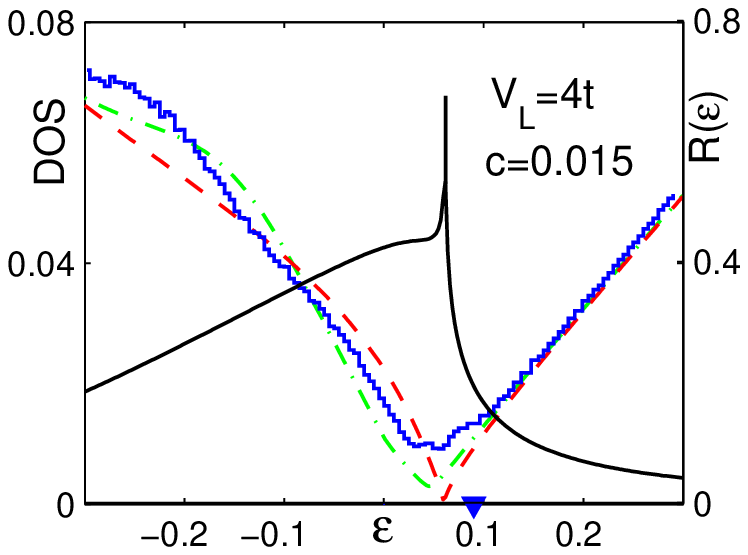}\\
\includegraphics[width=0.475\textwidth]{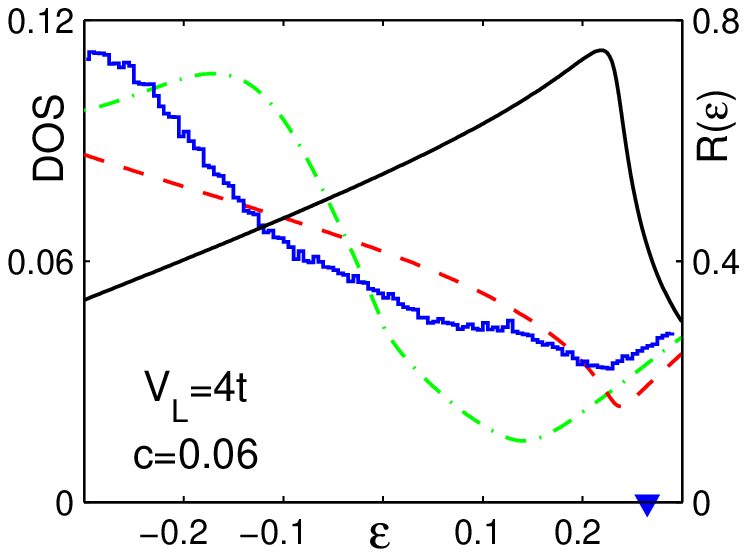}
\end{tabular}
\caption{(color online) A set of figures corresponding to impurity
perturbation $V_L=4t$ and different concentrations. Critical
concentration is  $c_r \approx 0.015$. Stepped curve stands for the
numerical computation, dashed --- the CPA, dash--dotted --- the ATA
(left Y-axis represents their values). Solid black curve is
$R(\varepsilon)$ (right Y-axis represents its values). Triangle on the
energy axis denotes the Fermi level position.} \label{DOS4t}
\end{figure}

\begin{figure*}
\begin{tabular}{cc}
\includegraphics[width=0.475\textwidth]{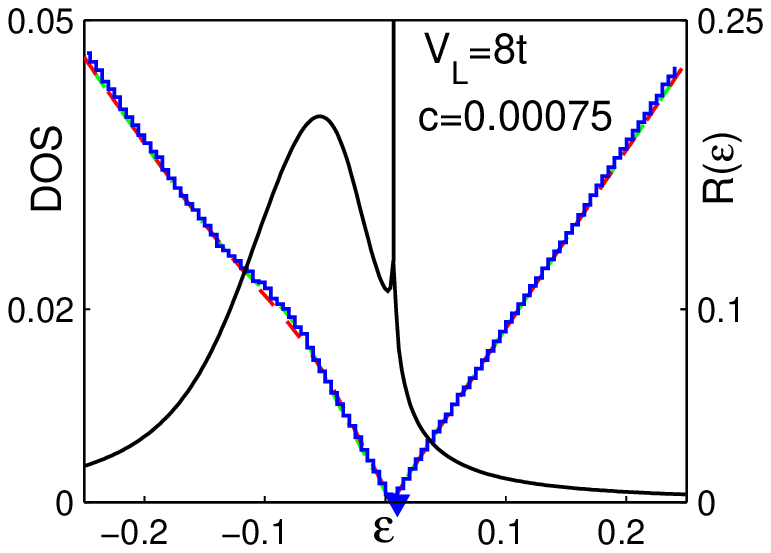}&%
\includegraphics[width=0.475\textwidth]{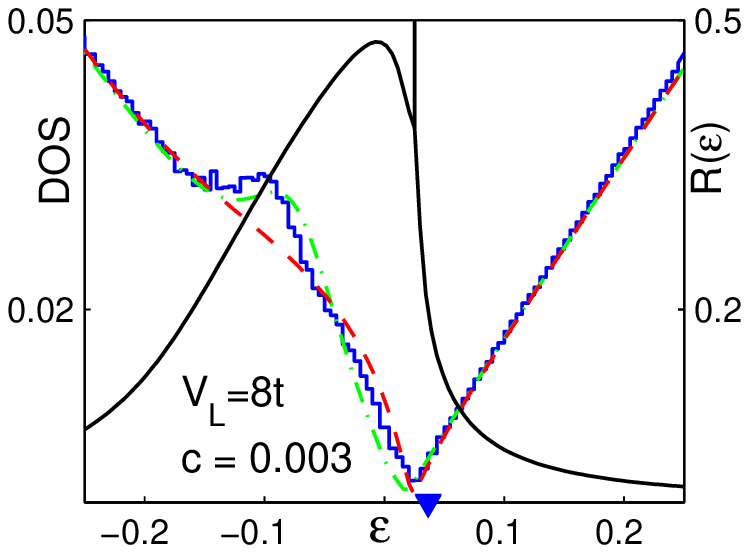}\\
\includegraphics[width=0.475\textwidth]{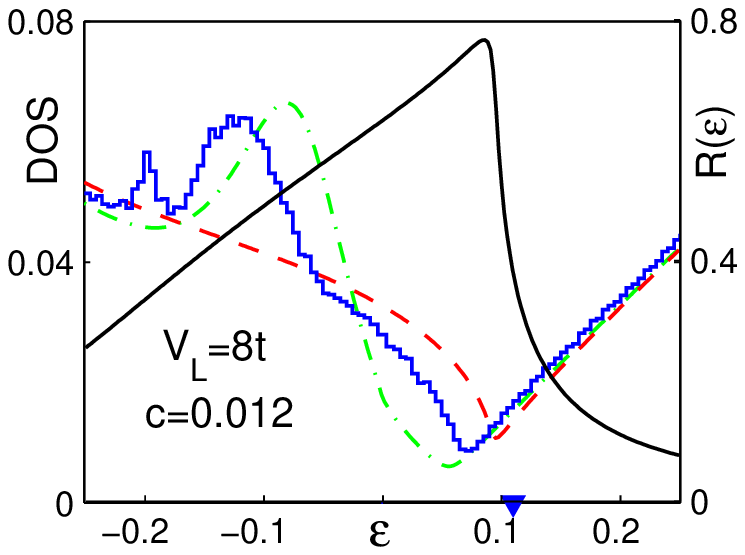}&%
\includegraphics[width=0.475\textwidth]{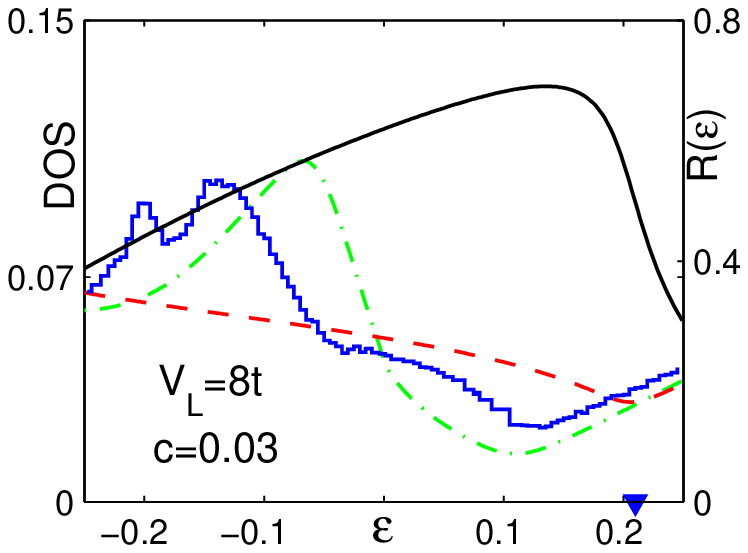}
\end{tabular}
\caption{(color online) A set of figures corresponding to impurity
perturbation $V_L=8t$ and different concentrations. Critical
concentration is  $c_r \approx 0.003$. Stepped curve stands for the
numerical computation, Dashed --- the CPA, dash--dotted --- the ATA
(left Y-axis represents their values). Solid black curve is
$R(\varepsilon)$ (right Y-axis represents its values). Triangle on the
energy axis denotes the Fermi level position.} \label{DOS8t}
\end{figure*}

\subsection{Numerical method}
Numerical techniques involving diagonalization of the random matrix
are too resource consuming to simulate disordered systems approaching
in their dimensions real experimental samples. However, information on
eigenvectors is superficial for the DOS calculations. So far, the
Haydock method,\cite{Haydock} based on an expansion of the diagonal
element of the Green's function into an infinite fraction, has been
extensively used for numerical calculation of the graphene
DOS.\cite{Pereira,Wu} Still, this approach is not without its
shortcomings. It is the local DOS (LDOS) that is calculated within this
approach. The necessity to truncate the infinite fraction at some
point sometimes leads to unphysical oscillations of the LDOS, which
are difficult to keep under control and to distinguish from actual
features of the spectrum. Furthermore, the total DOS is obtained in
the Haydock method by averaging the LDOS at several lattice sites, and
an inclusion of all sites in the model system into the averaging
process is absolutely impractical. The above leaves a touch of
uncertainty in the DOS minutiae. In a contrast, we relied  on the
Dean's calculation scheme.\cite{Dean} It allows to obtain the total
number of eigenvalues of a hermitian matrix that are less than a
specified value. This provides a possibility to explore the DOS with a
desired degree of precision and to preserve all particularities of the
resulting curve.

This method has proven to be especially effective for 1D systems. The
time required to finish a single Dean's algorithm loop is proportional
to the number of atoms ($N$) in a 1D system, which is fast enough to
simulate really large 1D systems. However, with an increase in the
system's dimensionality, the computational time required for one loop
increases. For a 2D system it is proportional to $N^2$.\cite{Dean} In
our case of graphene, we obtained DOSs for the system comprised of
$5.3\cdot 10^6$ atoms, which corresponds to a system with the linear
dimensions about $0.3\mu m$ --- about the size of real experimental
samples. To eliminate the influence of boundary states on the DOS we
applied periodical boundary condition for the zigzag boundary of the
model system under consideration.\cite{zigzag} The numerically
calculated DOS for the described model system is given in
Fig.~\ref{ideal}. Some jaggedness seen in the DOS curve is related to
the finite size of the model system. 

\begin{figure*}
\begin{tabular}{cc}
\includegraphics[width=0.475\textwidth]{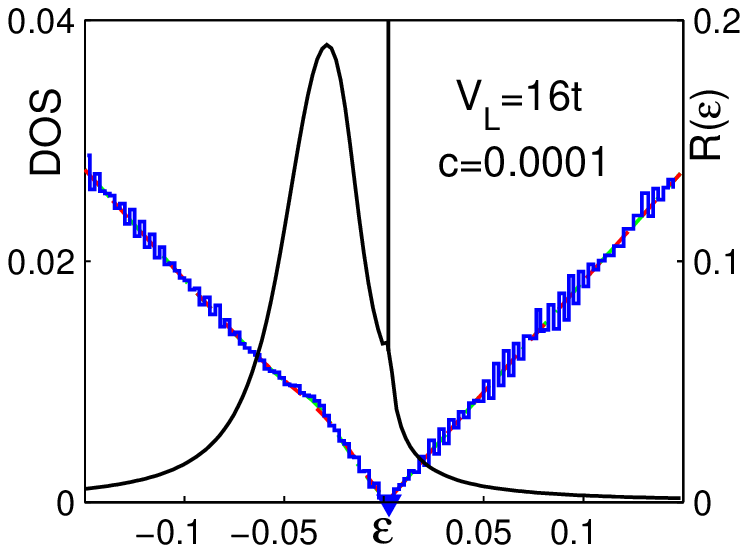}&%
\includegraphics[width=0.475\textwidth]{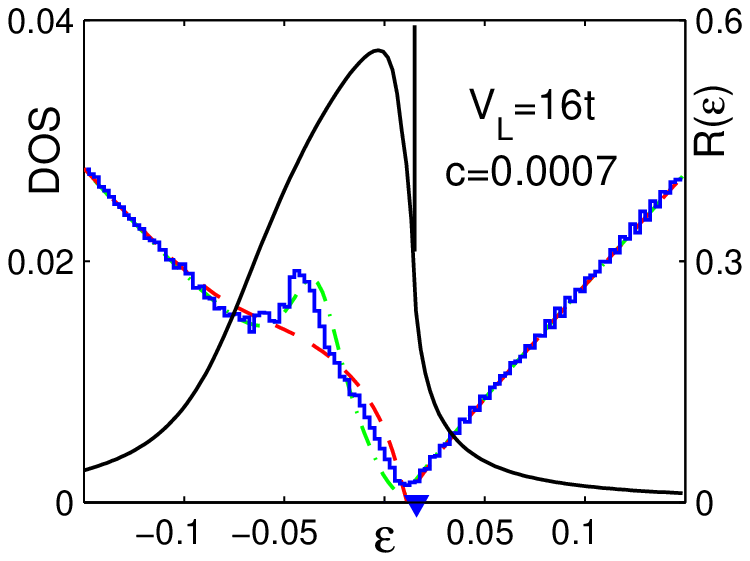}\\
\includegraphics[width=0.475\textwidth]{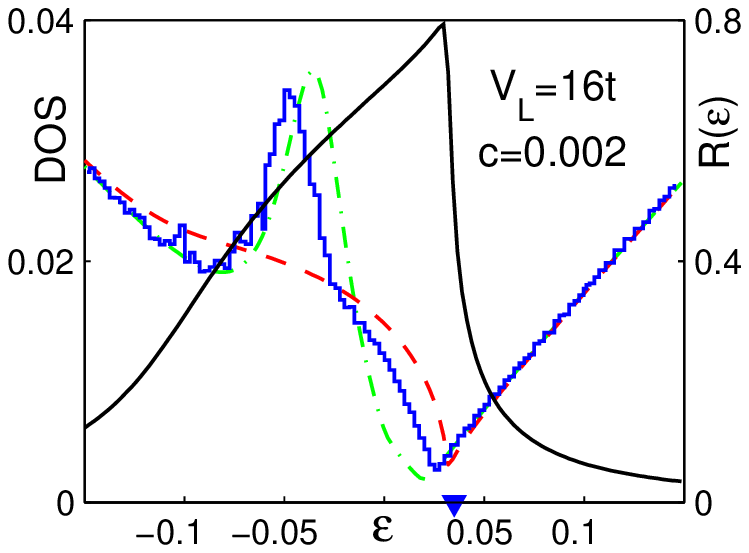}&%
\includegraphics[width=0.475\textwidth]{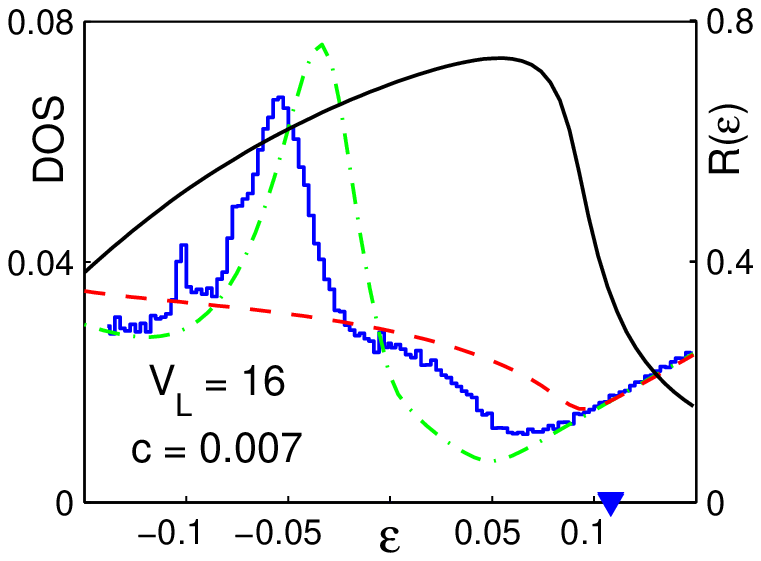}
\end{tabular}
\caption{(color online) A set of figures corresponding to the impurity
perturbation $V_L=16t$ and different concentrations. Critical
concentration is  $c_r \approx 0.0007$. Stepped curve stands for the
numerical computation, dashed --- the CPA, dash--dotted --- the ATA
(left Y-axis represents their values). Solid black curve is
$R(\varepsilon)$ (right Y-axis represents its values). Triangle on the
energy axis denotes the Fermi level position.} \label{DOS16t}
\end{figure*}

\section{Results and discussion}

The figures Fig.~\ref{DOS4t}, Fig.~\ref{DOS8t}, and Fig.~\ref{DOS16t}
correspond to impurity perturbations $V_L=4t$, $V_L=8t$, and
$V_L=16t$, respectively. At the negative impurity potentials $V_L$ the
whole picture is simply mirrored against the zero energy. For each
perturbation magnitude we consider qualitatively different regions of
impurity concentration: $c<c_r$ (before the spectrum rearrangement),
$c\approx c_r$ (in the course of the spectrum rearrangement) and
$c>c_r$ (after the spectrum rearrangement). We plot the CPA DOS,
the ATA DOS, and the numerically calculated DOS, with the left Y-axis
representing their values. We add the CPA applicability criterion by
plotting the small parameter of the series $R(\varepsilon)$ (solid
line), with the right Y-axis showing its values in the same
figures. We also designated by the triangle the Fermi level position,
obtained from the numerical data for the impure system.  

For the low concentrations (i.e. those that are less than $c_r$),
analytical curves, namely the CPA DOS and the ATA DOS, perfectly fit
the numerical histogram. The DOS only slightly deviates from the
conventional Dirac DOS mainly because of the shift towards positive
energies. The applicability criterion $R(\varepsilon)<0.5$ is
satisfied practically at all energies within the chosen window,
$R(\varepsilon)$ is small and characteristically contains two
maxima. The sharp one corresponds to the van Hove singularity and
predicts failure of analytical approximations in the Dirac point
vicinity. Less sharp one is due to the $\sigma^2$ factor.  

When the impurity concentration is increased approximately to the
critical value $c_r$, maxima of small parameter $R(\varepsilon)$ show
the tendency to merge together into a single maximum, which height goes
beyond the $0.5$ value (as it was shown in
Ref.~\onlinecite{YuV}). This event indicates the onset of the spectrum
rearrangement, providing a reference point for the critical
concentration at the given $V_L$. In the domain with heightened values
of $R(\varepsilon)$ the discrepancy between the CPA DOS and the
simulation results is more clearly expressed.

Figures also show that the perturbation $V_L=4t$ is marginal as the
resonance state appears at the periphery of the region, where the
Dirac approximation (\ref{gf0}) works well. Even for low
concentrations some divergence is seen between analytical and numerical
curves at the edges of energy domain considered. Impurity resonances
are smeared out and, therefore, cannot be readily discerned for a
perturbation of this strength. Likewise, for such a $V_L$ the impurity
resonance is not well defined in the single impurity LDOS. It should
be mentioned that in the weak scattering regime ($V_L<W$) the spectrum
rearrangement process does not take place at all (as it was evident in
Ref.~\onlinecite{Pereira}, when the average DOS maintained the mere
rigid shift with increasing impurity concentration).

With an increase in $V_L$ the impurity resonance becomes well
defined. It has been obvious beforehand that the CPA DOS should not
contain any sharp features in a contrast to the ATA DOS. Because of
the absence of self--consistency  the ATA gravitates more to the
single--impurity resonance. It is clearly seen in Fig.~\ref{DOS8t} and
Fig.~\ref{DOS16t} that the ATA DOS quite correctly reproduces
the resonance peak at the impurity concentrations that are
close to the critical one. The larger is the impurity potential $V_L$,
the better the ATA curve fits numerical data. 

This coincidence, however, is not pertained to the impurity
concentrations that exceed the critical concentration of the spectrum
rearrangement. With an increase in the impurity concentration the
maximum position shifts in the direction of negative energies from the
energy of the single--impurity resonance and the second, accompanying
maximum is coming forth. While the shift of the primary maximum from
the Lifshitz equation root is considerable, the maximum in the ATA DOS
remains still at the single--impurity resonance energy. The
aforementioned irregular structure in the DOS is the most intriguing
feature of the spectrum which cannot now be interpreted with the help
of the available CPA or ATA approximations. The maxima in the DOS are
located within a large domain, in which the CPA DOS does not follow
the simulated curve and, correspondingly, the CPA validity criterion
is not satisfied ($R(\varepsilon)>0.5)$. This domain covers the
single--impurity resonance and the minimum in the DOS, at which
valence and conduction bands are docking each other.    

\begin{figure}
\includegraphics[width=0.475\textwidth]{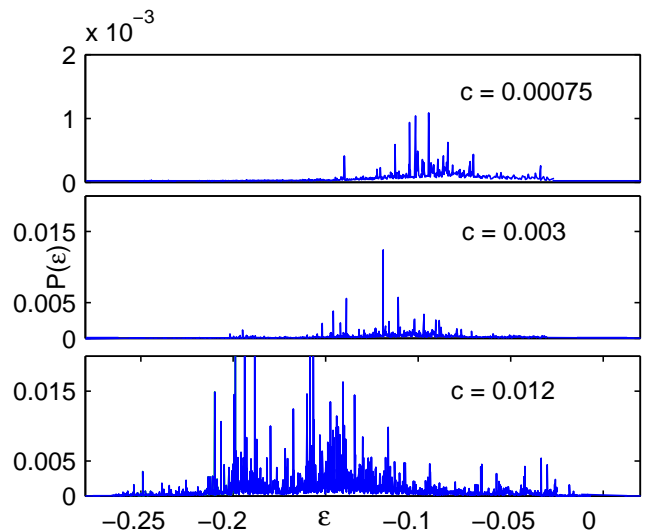}
\caption{(color online) Inverse participation ratio for $V_L=8t$ and
different impurity concentrations.}
\label{ipr}
\end{figure}

\begin{figure}
\begin{tabular}{c}
\includegraphics[width=0.475\textwidth]{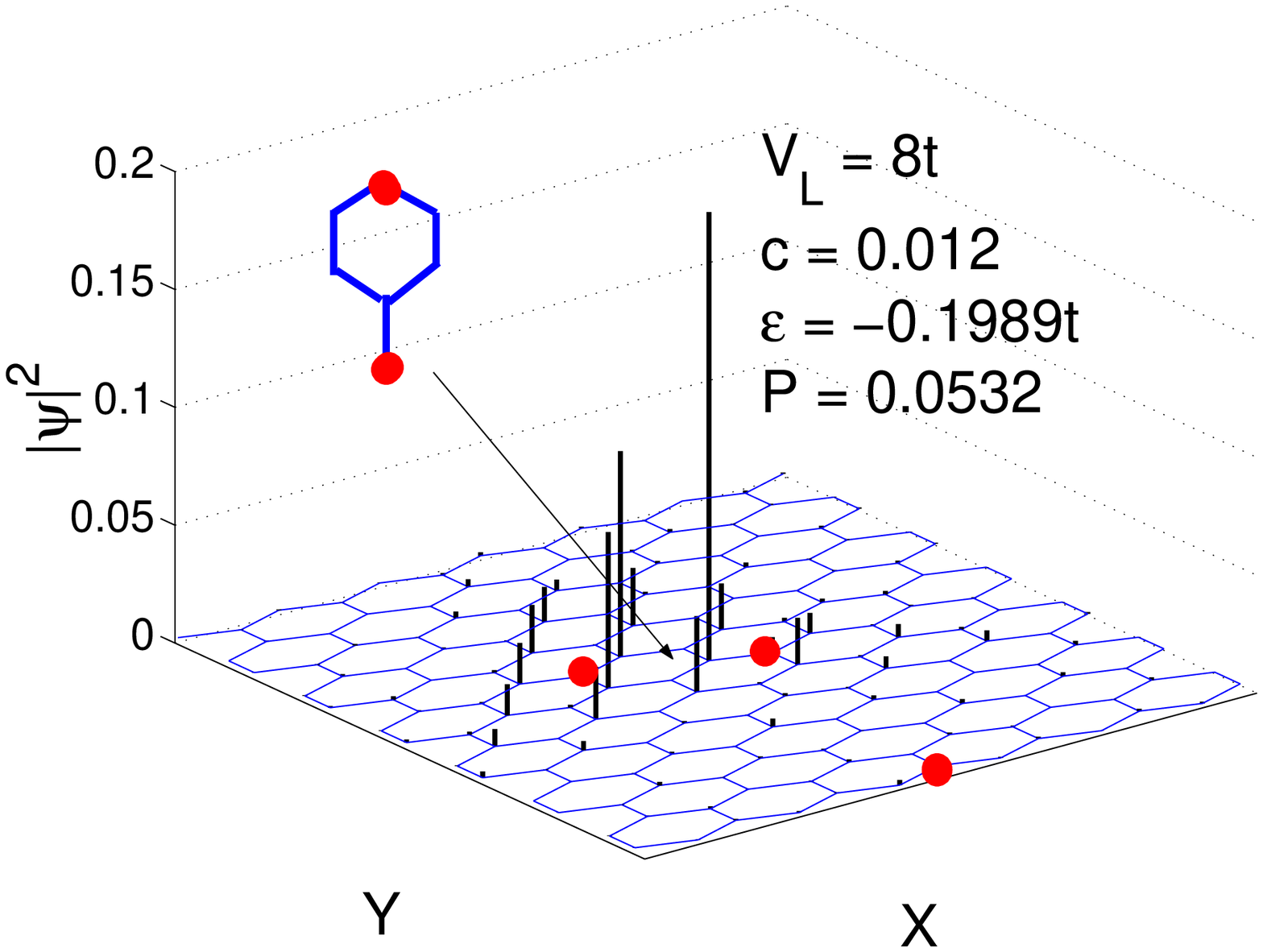}\\
\includegraphics[width=0.475\textwidth]{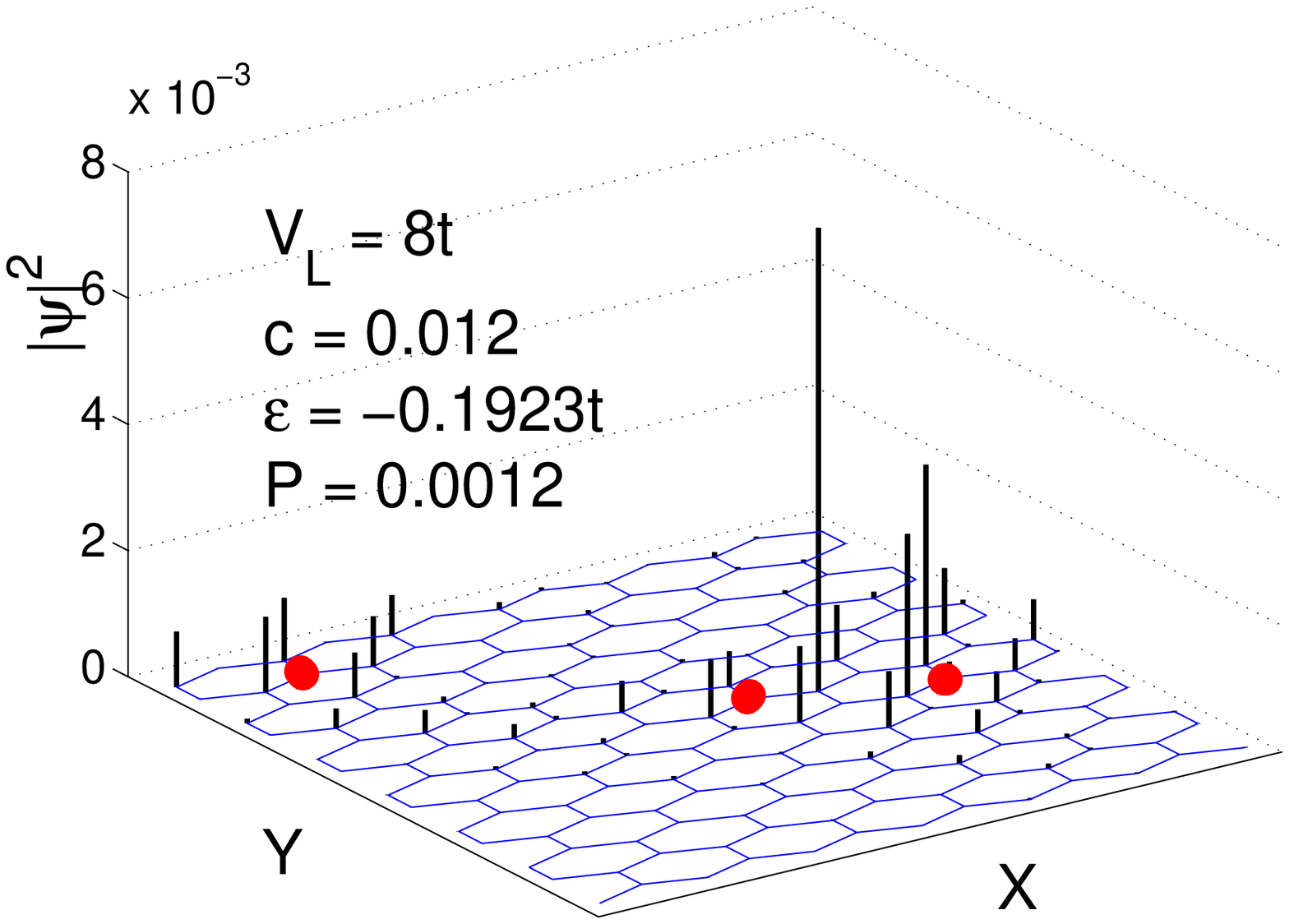}
\end{tabular}
\caption{(color online) Fragments of eigenstates for $V_L= 8t$ and $c
= 0.012$ are depicted. These configurations of impurities are
characteristic for the second peak in the DOS.} \label{IPR1}
\end{figure}

In addition, to study the character of states within this domain, we
calculated the inverse participation ratio\cite{Pereira}
$P(\varepsilon)$ as a localization criterion in a system of a smaller
size: 
\begin{equation}
P(\varepsilon) = \sum\limits_{\bm{n}_\alpha}|\psi_{\bm{n}_\alpha}|^4,
\end{equation}
where summation runs over all lattice sites. Even though the
comparison for systems of different size is not included in the
current article, we should mention that the hight of $P(\varepsilon)$
curves does not diminish with increasing the size of the system
suggesting electron localization. The dynamics of the inverse
participation ratio with increasing the impurity concentration is
given in Fig.~\ref{ipr}. Here again periodical boundary conditions
were used at zigzag edges of the sample to get rid of the sharp peak
at $\varepsilon = 0$, which is related to boundary states. Chosen
concentrations repeat those from Fig.~\ref{DOS8t} that corresponds
to the same impurity potential. A radical localization intensification
after the spectrum rearrangement is evidently seen. States are showing
a tendency for their localization in the very region, in which the CPA
is not valid. This fact also confirms a close connection between the
CPA applicability criterion and the Ioffe-Regel criterion.\cite{Ioffe}
When comparing Fig.~\ref{ipr} to Fig.~\ref{DOS8t}, in particular at
$c=0.012$, it is obvious that the largest values of $P(\varepsilon)$
match the peaks in the DOS. 

\begin{figure}
\begin{tabular}{c}
\includegraphics[width=0.475\textwidth]{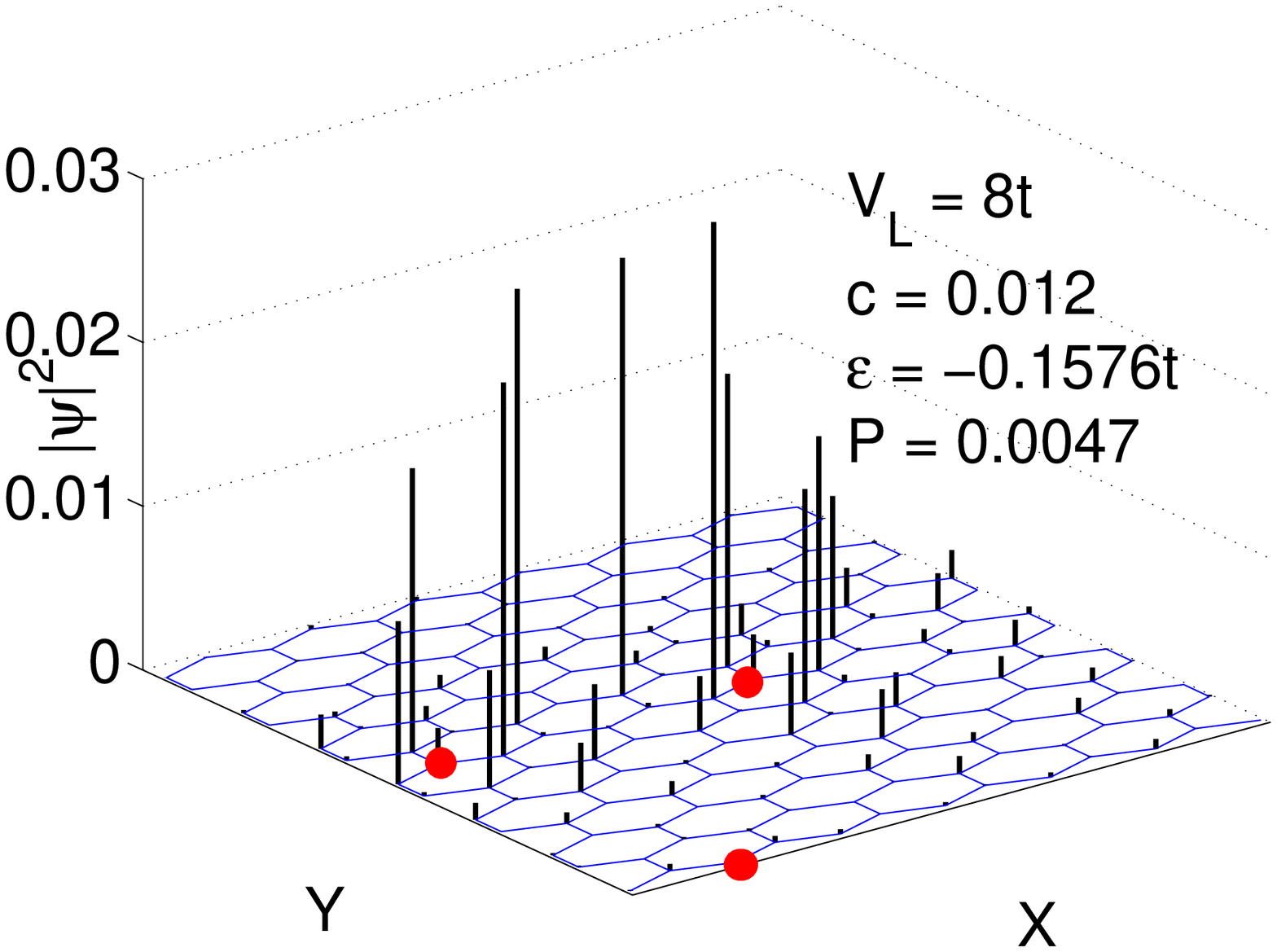}\\
\includegraphics[width=0.475\textwidth]{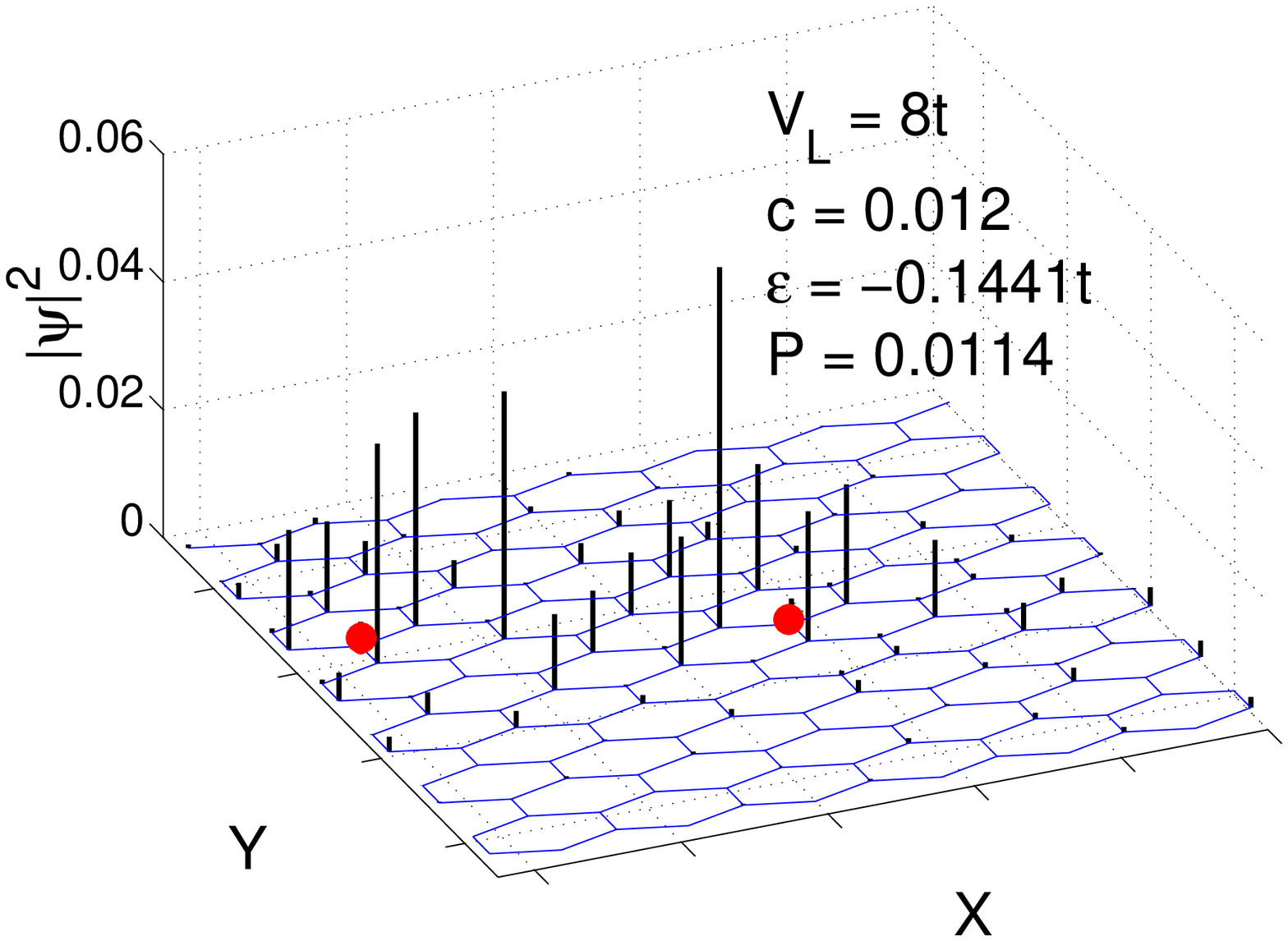}
\end{tabular}
\caption{(color online) Fragments of the eigenstates for $V_L= 8t$ and
$c = 0.012$ are depicted. They correspond to the first (closest to the
Dirac point) peak in the DOS.} \label{IPR2}
\end{figure}

A pair of states corresponding to the sharp peaks in the inverse
participation ratio graph are shown in Figs.~\ref{IPR1} and
\ref{IPR2}. The states corresponding to the first (counting from
$\varepsilon = 0$) maximum in the DOS at $\varepsilon \approx -0.14t$
are mostly represented by relatively distant impurity pairs and
triads. Equally challenging is the origin of the second DOS peak at
$\varepsilon \approx -0.19t$. The visualization of the wave-function
belonging to this region is provided in Fig.~\ref{IPR1}. It shows that
this peak is largely due to the characteristic pattern of impurities,
which is depicted in the same figure. It is worth mentioning that
impurity atoms are located on one sublattice for these strongly
localized states, while the $\psi$-function is concentrated on the
other sublattice. It resembles the situation with a double
impurity\cite{Katsnelson} and can be attributed to the relation
$|g_{\bm{0}_\alpha \bm{n}_\alpha}|\ll |g_{\bm{0}_\alpha%
\bm{n}_\beta}|$ for $|\varepsilon| \rightarrow 0$. To summarize,
when the critical concentration of impurities is exceeded, a quasigap
filled with localized states is developing in the graphene's spectrum
because of the impurity cluster effects. Since the CPA does not
account for cluster effects and scatterings by impurity clusters
dominate within this quasigap, the CPA is not applicable in this
region. 

Numerical results show that Dirac point as such is eliminated from the
spectrum when the critical concentration $c_r$ is
reached. Consequently, it is not justifiable to speak about the Dirac
point existence, albeit the impurity concentration can be relatively
low (as low as $c_r$ is). The CPA and the ATA do not correctly
describe the DOS minimum between the valence band and the conduction
band for $c>c_r$. Normally, the Dirac point coincides with the Fermi
energy for the pure (or undoped) graphene. However, at the finite
concentration of impurities situation changes drastically. Dean's
approach allows to track the position of the Fermi level, since it
outputs the total number of states located below any given energy. The
Fermi level monitoring revealed that its position shifts in the
positive direction away from the DOS minimum. The greater is the
impurity concentration the more prominent is the Fermi energy shift
from the DOS minimum, which transforms graphene into a ``doped''
conductor without any gate voltage applied. Should impurity atoms
bring additional electrons in the system, the doping effect will be
increased. 

This shift can be explained by the following uncomplicated
consideration. Without the conduction band, the valence band should
develop a strong tail above its upper edge for a positive impurity
potential $V_L>0$. Since the conduction band is not separated from the
valence band, this tail falls inside the conduction band producing
excessive states above the DOS minimum. Keeping in mind that total
number of states within the single band should be preserved because of
the sum rules, this expansion of the valence band into the conduction
band yields a deficit of states below the DOS minimum. At the constant
amount of carriers, extra electrons are flowing out to the conduction
band altering the Fermi level position. With increasing impurity
concentration the tail is gradually becoming more pronounced, which
consequently makes the shift in the Fermi energy more apparent.

\section{Conclusion}
A comprehensive analysis of the spectrum rearrangement in graphene
with substitutional impurities by numerical simulation has been
carried out. We studied the DOS near the Fermi level in graphene for a
set of impurity potentials and impurity concentrations.  

It was demonstrated that indeed a certain characteristic concentration
of impurities can be specified, at which the graphene's spectrum
undergoes a qualitative change. This critical impurity concentration
is associated with the spatial overlap of individual impurity
states. In a turn, it has been established that the cardinal
modification of the spectrum is manifested by the opening of the
filled with highly localized states quasigap around the impurity
resonance energy. The cluster effects were found to be responsible for
the quasigap formation. Pair impurity states representing the most
prominent peaks in the DOS within the quasigap have been visualized,
which emphasized the dominance of scatterings on impurity clusters
inside the quasigap. Aforesaid confirmed the predicted scenario of the
spectrum rearrangement in graphene.

A comparison of the CPA DOS with the numerically simulated DOS
supported the suggested CPA applicability criterion and its efficacy
as an instrument in the description of the spectrum rearrangement
passage. As well, intimate correlation between the CPA validity and
the degree of electron localization has been revealed. That is, inside
the quasigap, in which cluster effects are essential and states are
localized, the CPA is not reliable. 

Furthermore, we report about the phenomenon of the Fermi level shift
from the DOS minimum --- a kind of a self-doping, which alters the
conductivity of impure graphene without gate voltage variation even in
the case of neutral impurities.

\begin{acknowledgments}
We are grateful to Prof. Yu.~G.~Pogorelov for useful and stimulating
discussions. This work was partially supported by the Special Program
of the Department of Physics and Astronomy and by the Scientific
Program ``Nanostructural Systems, Nanomaterials and Nanotechnologies''
(grant No. 10/07-N) of the National Academy of Sciences of Ukraine. 
\end{acknowledgments}

\end{document}